\documentclass[twocolumn,showpacs,preprintnumbers,amsmath,amssymb]{revtex4-1}

\usepackage{graphicx}
\usepackage{dcolumn}
\usepackage{bm}
\usepackage{soul, xcolor}
\usepackage{color, wasysym}
\usepackage{textcomp}
\usepackage{amsmath, amssymb, amsthm, latexsym, epsfig}

\DeclareMathOperator{\D}{d\!}
\DeclareMathOperator{\E}{e}

\def\ulamek#1#2{\mbox{\normalfont$\frac{#1}{#2}$}}

\begin{document}

\newtheorem{theorem}{Theorem}
\newtheorem{definition}{Definition}
\newtheorem{lemma}{Lemma}
\newtheorem{proposition}{Proposition}
\newtheorem{remark}{Remark}
\newtheorem{con}{Conjecture}
\newtheorem{example}{Example}


\title{Integral decomposition for the solutions of the generalized Cattaneo equation}

\author{K. G\'{o}rska}
\email{katarzyna.gorska@ifj.edu.pl}
\affiliation{Institute of Nuclear Physics, Polish Academy of Sciences, \\ ul. Radzikowskiego 152, PL-31342 Krak\'{o}w, Poland}


\begin{abstract}
We present the integral decomposition for the fundamental solution of the generalized Cattaneo equation with both time derivatives smeared through convoluting them with some memory kernels. For power-law kernels $t^{-\alpha}$, $\alpha\in(0,1]$ this equation becomes the time fractional one governed by the Caputo derivatives which highest order is 2. To invert the solutions from the Fourier-Laplace domain to the space-time domain we use analytic methods based on the Efross theorem and find out that solutions looked for are represented by integral decompositions which tangle the fundamental solution of the standard Cattaneo equation with non-negative and normalizable functions being uniquely dependent on the memory kernels. Furthermore, the use of methodology arising from the theory of complete Bernstein functions allows us to assign such constructed integral decompositions the interpretation of subordination. This fact is preserved in two limit cases built into the generalized Cattaneo equations, i.e., either the diffusion or the wave equations. We point out that applying the Efross theorem enables us to go beyond the standard approach which usually leads to the integral decompositions involving the Gaussian distribution describing the Brownian motion. Our approach clarifies puzzling situation which takes place for the power-law kernels $t^{-\alpha}$ for which the subordination based on the Brownian motion does not work if $\alpha\in(1/2,1]$.  
\end{abstract}

                             

\maketitle

\section{Introduction}

Diffusion influences the course of kinetic phenomena investigated in many branches of pure physics, chemistry, and biology but its applications are by no means restricted to the fundamental problems of Nature. Knowledge of the origin and effects of diffusion mechanisms is necessary to understand diversity of results obtained in life, earth and environment sciences as well as in materials engineering and technological processes. Diffusion governs behaviour of phenomena occurring both in simple and complex systems, from the Brownian motion discussed in the course of the school physics to very special and advanced problems, like the transport phenomena in tissues and cells {\cite{dif1,JHJeon11,JFReverey15}}, spreading infectious diseases and drug administration \cite{dif2}, using zeolites in  catalysis \cite{dif3}, environmental science \cite{dif4}, materials engineering  and manufacturing new materials \cite{dif5,diff5a}, and even finances \cite{diff5b}. Comprehensive information on a diffusion process under study is not easy to get. Usually it strongly depends on the properties of the system under consideration and routinely its quantitive analyses start from determining the relation between the time $t$ and the mean square displacement (MSD) $\langle x^{2}(t) \rangle$ of diffusing objects. For the case of Brownian motion it takes the form of the Einstein-Smoluchowski relation $\langle x^{2}(t) \rangle \propto t$. The latter is valid for any $t\in\mathbb{R}_{+}$ and besides of the Brownian motion characterizes also other transport processes classified, in general, as the normal diffusion. Processes which exhibit deviations from the Einstein-Smoluchowski relation, e.g., lead to the relation $\langle x^{2}(t) \rangle \propto t^{\mu}, \mu>0$ (valid at least asymptotically for small and/or large time) {\cite{JWHaus87,RMetzler14,RMetzler19}},  are classified  as the anomalous diffusion and are nick-named the sub- and superdiffusion in dependence of the value of $\mu$. For $\mu \in (0, 1)$ we deal with the subdiffusion whereas the cases $\mu > 1$ are called the superdiffusion. If $\mu=2$ then the process bears the special name of ballistic motion. Here it has to be noted that contrary to the Einstein-Smoluchowski relation supposed to be satisfied for all admissible range of time  there exist experimentally observed relations between  $\langle x^{2}(t) \rangle$ and $t$ which change their functional shape as the time flows on. Retrieving such information provides us with some hints how to choose the type of evolution equation suitable to describe correctly the process under investigation. Well-researched approach is to modify the standard diffusion equation into its fractional counterpart             
\begin{equation}\label{22/03-1}
\partial_{t} p(x,t) = B \partial_{x}^{2} p(x, t) \longrightarrow 
{_{c}D}^{\mu}_{t}p(x,t) = B \partial_{x}^{2} p(x, t),
\end{equation}
with ${_{c}D}^{\mu}_{t}$ being the (Caputo) fractional time derivative of the order $\mu$, which leads to $\langle x^{2}(t) \rangle \propto t^{\mu}$. This simple modification is not enough if the behavior of MSD changes its shape with time - in such a case one has to go beyond Eq. \eqref{22/03-1} and consider more complicated evolution equations like the standard or fractional Cattaneo equations comprising more than one time derivative term and possibly involving more general forms of the time smearing than the standard fractional derivatives do. Adequate example have been studied recently in \cite{KGorska20} and reads     
\begin{multline}\label{18/01-3}
\tau\int_{0}^{t} \eta(t - \xi) \partial_{\xi}^{2} q_{\hat{\gamma}}(\tau; x, \xi) \D\xi + \int_{0}^{t} \gamma(t - \xi) \partial_{\xi} q_{\hat{\gamma}}(\tau; x, \xi) \D\xi \\ = B\, \partial_{x}^{2} q_{\hat{\gamma}}(\tau; x, t)
\end{multline}
where $B$ is the diffusion coefficient. {In what follows the subscript '$\hat{\gamma}$' is used to emphasize the role played by the Laplace transformed function $\gamma(t)$, namely $\hat{\gamma}(z) = \mathcal{L}[\gamma(t); z]$. The solution $q_{\hat{\gamma}}(\tau; x, t)$, if it is to be interpreted as the probability density (PDF) of finding the particle in a position $x \in \mathbb{R}$ at an instant of time $t > 0$, must be  shown non-negative and normalizable for chosen memory functions $\eta(t)$ and $\gamma(t)$. The memory functions $\eta(t)$ and $\gamma(t)$ express the smearing of the second and first time derivatives, respectively, and for special choices reduce Eq. \eqref{18/01-3} to more familiar cases:  for $\eta(t)=0$ Eq. \eqref{18/01-3} becomes the time smeared diffusion equation
\begin{equation}\label{28/01-1}
\int_{0}^{t} \gamma(t - \xi) \partial_{\xi} N_{\hat{\gamma}}(x, \xi) \D\xi = B\, \partial_{x}^{2} N_{\hat{\gamma}}(x, t) 
\end{equation}
while for  $\gamma(t)=\eta(t)=\delta(t)$, with $\delta(t)$ being the Dirac distribution, Eq. \eqref{18/01-3} is the standard Cattaneo equation \cite{CRCattaneo48, CRCattaneo58}. The striking feature of the latter is that its fundamental  solution (i.e. evolution of the space localized time initial condition), similarly to the case of wave equation, does vanish outside the compact support $\Delta(x,t)\in\mathbb{R}\otimes\mathbb{R}_{+}$ \cite{JAStratton41,PMMorse53,JMasoliver96}. Thus it appears justified to ask if already mentioned properties, i.e., the functional dependence on $t$ which MSD obeys in $t$ and the compact character of $\Delta (x,t)$  are kept for the time smeared Cattaneo equation Eq. \eqref{18/01-3}. To judge the problem we shall consider the power-law memory functions being the most frequently used type of functions adopted to mimic the memory effects.}

Looking for the answer to the above questions let us note that Eqs. \eqref{18/01-3} and \eqref{28/01-1} differ only by the term containing the parameter $\tau$. For small $\tau$ Eq. \eqref{18/01-3} formally tends to Eq. \eqref{28/01-1}, i.e., to the time smeared diffusion equation. On the opposite, for large $\tau$, the smeared second time derivative dominates and we can neglect the smeared first time derivative.  So we arrive at the so-called time-smeared diffusion-wave equation which, if detailed to the fractional derivative case, was studied in \cite{TSandev18, YuLuchko13, YuLuchko19}. Hence, Eq. \eqref{18/01-3}, with $\tau$ treated as a free parameter, describes all intermediate cases. This feature should be also reflected in the limit behaviour of $q_{\hat{\gamma}}(\tau; x, t)$: 
\begin{align}\label{20/04-1}
N_{\hat{\gamma}}(x, t)=\lim_{\tau\to 0}q_{\hat{\gamma}}(\tau; x,& t) \quad \text{and} \\ \label{20/04-2}
&W_{\hat{\gamma}}(x, t)=\lim_{\tau\to\infty} q_{\hat{\gamma}}(\tau; x, t),
\end{align}
where in the shorthand notation these two limits will be marked as $q_{\hat{\gamma}}(0; x, t)$ and $q_{\hat{\gamma}}(\infty; x, t)$.

The solutions $N_{\hat{\gamma}}(x, t)$ and $W_{\hat{\gamma}}(x, t)$ can be represented in the form of integral decomposition introduced in Ref. \cite{HCFogedby94}. The characteristic property of such a formalism  is that one replaces their physically observed dependence on the space coordinate $x$ and the physical (laboratory) time coordinate $t$ by the braided distributions of the space coordinate $x$ evolving according to an internal (operational) time $\xi$ which, in turn, is assumed to be stochastically dependent on the clock measured laboratory time $t$. Integral decomposition entwines both distributions, threated as independent, through the integral taken over internal, or operational, time $\xi$. From the widely known results, e.g, \cite{IMSokolov02, IMSokolov05, ABaule03, AChechkin21}, we learn that $N_{\hat{\gamma}}(x, t)$ can be given by 
\begin{equation}\label{8/02-10}
N_{\hat{\gamma}}(x, t) = \int_{0}^{\infty} N(x, \xi) f_{0}(\xi, t) \D\xi,
\end{equation}
where
\begin{equation}\label{3/03-2}
N(x, t) = \E^{-x^{2}/(4 B t)}/\sqrt{4 \pi B t}  
\end{equation}
and 
\begin{equation}\label{3/02-2}
f_{0}(\xi, t) = \mathcal{L}^{-1}[\hat{\gamma}(z) \E^{-\xi z \hat{\gamma}(z)}; t]. 
\end{equation}
As presented in  \cite[Eq. (11)]{TSandev18} for the case of the wave-diffusion equation and its solution $W_{\hat{\gamma}}(x, t)$ the integral decomposition has the form of Eq. \eqref{8/02-10} but instead of $f_{0}(\xi, t)$ one has 
\begin{equation}\label{10/03-1}
f_{\infty}(\xi, t) = \mathcal{L}^{-1}\{z \hat{\eta}(z) \exp[-\xi z^{2} \hat{\eta}(z)]; t\}.
\end{equation}
Note that  subscripts '$0$' and '$\infty$' in Eqs. \eqref{3/02-2} and \eqref{10/03-1} are used in the context of previously discussed limits of the generalized Cattaneo equation. Within the investigation of diffusion phenomena rooted in the stochastic processes approach it is said that $f_{0}(\xi, t)$ or $f_{\infty}(\xi, t)$ subordinate the normal distribution if all functions involved in the game: $N(x, \xi)$, $f_{0}(\xi, t)$ and $f_{\infty}(\xi, t)$ are independent PDFs given by non-negative, normalized, and infinitely divisible functions. These conditions are satisfied by the normal distribution $N(x, t)$. {Nevertheless, the problem arises with the functions $f_{0}(\xi, t)$ and $f_{\infty}(\xi, t)$ which may or may not obey all these conditions.  Consequently, it is needed to distinguish two classes of $f_{..}(\cdot,-)$'s - if they are independent PDFs satisfying all just mentioned requirements then due to \cite{IMSokolov02, AChechkin21} are named ``safe'' functions;  in the opposite case they are called ``dangerous''. Besides of being independent on $N(x, \xi)$ the sufficient condition for the functions $f_{0}(\xi, t)$ and $f_{\infty}(\xi, t)$ to be the ``safe'' ones is fulfilled} if $z \hat{\gamma}(z)$ or $z^{2} \hat{\eta}(z)$ belong, for $z$ restricted to $\mathbb{R}_{+}$, to the class of  the so-called completely Bernstein functions (CBF) which definition is quoted in the Appendix \ref{a1}. Concluding this part of our considerations: usage of subordination approach  works correctly if $f_{0}(\xi, t)$ or $f_{\infty}(\xi, t)$, respectively, are PDFs independent on $N(x, \xi)$ and are given by the completely Bernstein functions. It is important to fully clarify the problem because subordination is the cornerstone of the stochastic processes framework and is based on introducing the operational time $\xi(t)$. The later in fact the operational time means a time-like parametrization of the composite (parent) process $X(\xi(t))$ given by the Brownian motion with jumps stochastically distributed in the operational time $\xi(t)$.

Throughout  the paper we use the relation between memories $\eta(t)$ and $\gamma(t)$ which comes out, as a constraint, from consistency conditions put on the  smearing of the first and second time derivatives. These  relations have been derived in \cite{KGorska20} and  according to them if the memory function $\gamma(t)$ is made responsible for the time smearing of the first time derivative then 
\begin{equation}\label{29/01-1}
\eta(t) = \int_{0}^{t} \gamma(\xi) \gamma(t - \xi) \D\xi, \qquad \hat{\eta}(z) = \hat{\gamma}^{2}(z)
\end{equation}
represents the smearing of the second time derivative. Authors of many theoretical and experimental studies concerning termoelasticity \cite{YZPovstenko15}, anomalous transport in complex media \cite{ACompte97,EAwad20,AGiusti18}, chemical reactions \cite{TKosztolowicz14, KDLewandowska08} are used to model the memory functions by the power-law functions $\eta(t) \propto t^{1 - 2\alpha}$ and $\gamma(t) \propto t^{-\alpha}$, $\alpha\in(0, 1]$ which in the Laplace domain read $\hat{\eta}(z) = z^{2(\alpha - 1)}$ and $\hat{\gamma}(z) = z^{\alpha - 1}$. In such a case $f_{0}(\xi, t)$ is always ``safe'', i.e., correctly defined PDF but, as shown in Ref. \cite{TSandev18}, $f_{\infty}(\xi, t)$ is \underline{not}  ``safe''  PDF for $\alpha\in (1/2, 1]$. Thus, for $\alpha\in (1/2, 1]$, the integral decomposition of the fundamental solution to the generalized Cattaneo equation based on the normal distribution must not be treated any longer as a subordination. From the other side, as shown in Ref. \cite{KGorska20}, this solution is non-negative and normalizable. In what follows we will present the solution to this puzzle and propose an alternative integral decomposition which keeps $f_{0}(\xi, t)$ but instead of $N(x, \xi)$ it involves the fundamental solution of the Cattaneo equation if  $\alpha\in (1/2, 1]$.

The paper is organized as follows. In Sec. \ref{sec2} we study two ways to obtain the integral decompositions suitable for the generalized Cattaneo equation. Following the first, usually adopted, one we arrive at the product of the normal distribution and the function which is not always ``safe'' PDF. Within the alternative approach the use of Efross theorem allows one to find integral decomposition in which the normal distribution is replaced by the solution of the standard Cattaneo equation. We show that in such a case we get the integral which integrand is the product of functions being ``safe'' PDFs. Thus, our integral decomposition expresses the subordination approach. In Sec. \ref{sec3}, taking the appropriate limit of $\tau$, we present the passage from the integral decomposition introduced in the paper into the known examples of integral decomposition got for the solutions of the anomalous diffusion and diffusion-wave equations. In Sec. \ref{sec4} our construction of integral decompositions is illustrated on three examples characterized by important role played by the memory effects - either the strictly localized case or the power-law memory functions or mixture  of these  two. Results of the paper are {resumed and discussed in Sec. \ref{sec5} while Sec. \ref{sec6} contains concluding remarks concerning prospective applications of the Efross theorem based integral decompositions used as a tool to investigate both some fundamental problems as well as hot topics recently emerging in the anomalous diffusion.  The paper is completed by five} appendixes devoted to explanation of mathematics used throughout the text.

\section{Integral decomposition}\label{sec2}

It is an obvious statement that any integral decomposition relied on factorization of the space and time coordinates is not unique. Using integral decompositions  to describe anomalous diffusion and non-Debye relaxations is commonly justified by physical and mathematical considerations based on modeling these phenomena by stochastic processes underlying them \cite{AStanislavsky17,AStanislavsky19}. Integral decompositions represent joint probabilities and provide us with the correct answer  if the integrals under study are not only non-negative and normalizable on $\mathbb{R}_{+}$ but also result from composition of infinitely divisible distributions \cite{AChechkin21,Bochner}. Namely this condition allows to equip them with probabilistic interpretation and to use methods elaborated within the theory of stochastic processes, among them the subordination approach \cite{AChechkin21,Bochner}. In what follows we shall show how to construct, without recalling probabilistic tools, integral decompositions which represent non-negative normalizable solutions to diffusion equations of the type Eq. \eqref{18/01-3}.   

\subsection{Integral decomposition preserving the $N(x, \xi)$ distribution} 
 
Eq. \eqref{8/02-10} with Eqs. \eqref{3/02-2} or \eqref{10/03-1} put in are special cases of the solution to Eq. \eqref{18/01-3}. In the Fourier-Laplace domain it reads
\begin{equation}\label{2/02-3}
\hat{\tilde{q}}_{\hat{\gamma}}(\tau; k, z) = \frac{z^{-1} \hat{M}_{\hat{\gamma}}(z)}{\hat{M}_{\hat{\gamma}}(z) + B k^{2}},
\end{equation}
where 
\begin{equation}\label{23/02-2}
\hat{M}_{\hat{\gamma}}(z) = \tau [z \hat{\gamma}(z)]^{2} + z \hat{\gamma}(z).
\end{equation}
The symbols '$\sim$' and $k$ are related to the Fourier transform whereas '$\wedge$' and $z$ correspond to the Laplace transform. Using $b^{-1} = \int_{0}^{\infty} \exp(-b\xi) \D\xi$ to get rid of the denominator in Eq. \eqref{2/02-3} we are able to rewrite Eq. \eqref{2/02-3} as
\begin{equation}\label{23/02-3}
\hat{\tilde{q}}_{\hat{\gamma}}(\tau; k, z) = \int_{0}^{\infty} \E^{-\xi B k^{2}}\, z^{-1} \hat{M}_{\hat{\gamma}}(z) \E^{-\xi \hat{M}_{\hat{\gamma}}(z)} \D\xi.
\end{equation}
The term $\exp(-\xi B k^{2})$ corresponds, by the inverse Fourier transform, to the normal distribution $N(x, \xi)$, whereas taking the inverse Laplace transform of the $z$ dependent term we get 
\begin{equation}\label{2/02-5}
q_{\hat{\gamma}}(\tau; x, t) = \int_{0}^{\infty} N(x, \xi)\, f_{\tau}(\xi, t) \D\xi
\end{equation}
where 
\begin{equation}\label{3/02-1}
f_{\tau}(\xi, t) = \mathcal{L}^{-1}[z^{-1} \hat{M}_{\hat{\gamma}}(z) \E^{-\xi \hat{M}_{\hat{\gamma}}(z)}; t].
\end{equation}
The function $f_{\tau}(\xi, t)$ tends to $f_{0}(\xi, t)$ for $\tau \to 0$ and to $f_{\infty}(\xi, t)$ for $\tau\to\infty$. To show that $f_{\tau}(\xi, t)$ is ``safe'' PDF means to prove that it is normalized, non-negative, and infinitely divisible. Normalizability is easily seen, $\int_{0}^{\infty} f_{\tau}(\xi, t) \D\xi = \mathcal{L}^{-1}[z^{-1}; t] = 1$. Non-negativity of $f_{\tau}(\xi, t)$ comes out from the Bernstein theorem presented in Appendix \ref{a1}. Due to this theorem $f_{\tau}(\xi, t)$ is non-negative if $z^{-1}\hat{M}_{\hat{\gamma}}(z) \exp[-\xi \hat{M}_{\hat{\gamma}}(z)]$, $z\in\mathbb{R}_{+}$, is given by a completely monotone function (CMF), i.e. non-negative function whose all derivatives exist on $\mathbb{R}_{+}$ and alternate \cite{RLSchilling10}. Because the product of CMFs is CMF then the Laplace transform of $f_{\tau}(\xi, t)$ is CMF if $z^{-1}\hat{M}_{\hat{\gamma}}(z)$ and $\exp[-\xi \hat{M}_{\hat{\gamma}}(z)]$ are CMFs. Both of these conditions are ensured for $\hat{M}_{\hat{\gamma}}(z)$ being CBF. The completely Bernstein character of $\hat{M}_{\hat{\gamma}}(z)$ guarantees also that $f_{\tau}(\xi, t)$ is an infinitely divisible distribution, see Appendix \ref{a2}. Thus, we end up with the conclusion that if $\hat{M}_{\hat{\gamma}}(z)$ is CBF then $f_{\tau}(\xi, t)$ is infinitely divisible PDF so it may be claimed that it subordinates $N(x, \xi)$.

Let us check if the above statement  appears conclusive for some special cases of Eq. \eqref{18/01-3}.  For the standard Cattaneo equation with $\hat{\eta}(z) = \hat{\gamma}(z) = 1$ we have $\hat{M}_{1}(z) = \tau z^{2} + z$ which is not CBF. So the appropriate $f_{\tau}(\xi, t)$ in not PDF and it does not subordinate $N(x, \xi)$. As the next example we take Eq. \eqref{18/01-3} with the power-law memory functions. Now $\hat{M}_{\hat{\gamma}}(z) = \tau z^{2\alpha} + z^{\alpha}$ is CBF only if $\alpha\in (0, 1/2]$ whereas for the remaining values of $\alpha$, that is $\alpha\in(1/2, 1]$, it is not CBF. It means that for this example $f_{\tau}(\xi, t)$ is a ``safe'' PDF only for $\alpha\in (0, 1/2]$ and it is ``dangerous'' if $\alpha\in(1/2, 1]$. Thus, it exhibits the behavior of $f_{\infty}(\xi, t)$ presented in \cite{TSandev18}. 

The above examples show that the integral decomposition given by Eq. \eqref{2/02-5} must not be always treated as the subordination of two processes. We shall propose another type integral decomposition within which, still keeping  $f_{0}(\xi, t)$ present, we will use another parent process instead of Gaussian distribution characteristic for the Brownian motion. 

\subsection{Integral decomposition preserving $f_{0}(\xi, t)$}\label{ssec2B}

To obtain the integral decomposition different than Eq. \eqref{2/02-5} we first calculate the Fourier transform of Eq. \eqref{2/02-3}. That gives
\begin{equation}\label{1/02-1}
\hat{q}_{\hat{\gamma}}(\tau; x, z) = \frac{z^{-1} [\hat{M}_{\hat{\gamma}}(z)]^{1/2}}{2 \sqrt{B}} \exp\Big\{- \frac{|x|}{\sqrt{B}} [\hat{M}_{\hat{\gamma}}(z)]^{1/2}\Big\},
\end{equation}
which can be rewritten as 
\begin{equation}\label{2/02-1}
\hat{q}_{\hat{\gamma}}(\tau; x, z) = \hat{\gamma}(z)  F(x,g(z)),
\end{equation}
where
\begin{equation}\label{23/02-5}
F(x,g) = \frac{1}{2 a} \frac{1/(2\tau) + g}{\sqrt{g^{2} - 1/(4\tau^{2})}} \E^{-\frac{|x|}{a} \sqrt{g^{2} - 1/(4\tau^{2})}}
\end{equation}
and  $g$ is a shorthand of $g(z)$ for simplifying the notation. The non-negative constant $a$ is equal to $(B/\tau)^{1/2}$ while the auxiliary function $g(z)$ reads
\begin{equation}\label{23/02-4}
g(z) = z \hat{\gamma}(z) + 1/(2\tau).
\end{equation}

Using the Efross theorem \cite{AEfross35,LWlodarski52,UGraf04,KGorska12a,AApelblat21} enables us to calculate the inverse Laplace transform of Eq. \eqref{2/02-1}. Here, we only sketch the derivation leaving its details  to Appendix \ref{a3}. According to the Efross theorem Eq. \eqref{2/02-1} can be written as
\begin{align}\label{3/02-5}
q_{\hat{\gamma}}(&\tau; x, t) = \int_{0}^{\infty}\!\! \mathcal{L}^{-1}[F(x,g); \xi] \mathcal{L}^{-1}[\hat{\gamma}(z) \E^{-\xi g(z)}; t] \D\xi \nonumber\\
& = \int_{0}^{\infty}\!\! \mathcal{L}^{-1}[F(x,g); \xi] \E^{-\xi/(2\tau)} \mathcal{L}^{-1}[\hat{\gamma}(z) \E^{-\xi z \hat{\gamma}}; t] \D\xi \nonumber \\
& = \int_{0}^{\infty}\!\! \mathcal{L}^{-1}[F(x,g); \xi] \E^{-\xi/(2\tau)} f_{0}(\xi, t) \D\xi,
\end{align}
where $f_{0}(\xi, t)$ is given by Eq. \eqref{3/02-2}. The product of $\mathcal{L}^{-1}[F(x,g); \xi]$ and $\exp[-\xi/(2\tau)]$ leads to the solution of the standard Cattaneo equation $q_{\rm C}(\tau; x, t)$. Namely, from the properties of the Laplace transform given by Eq. \eqref{20/04-3} we write down this product as $\mathcal{L}^{-1}[F(x,g + 1/(2\tau)); \xi]$. Hence, it can be expressed as 
\begin{equation}\label{23/02-6}
\frac{1}{2a} \frac{1/\tau + g}{\sqrt{g^{2} + g/\tau}} \exp\Big(\!-\frac{|x|}{a}\sqrt{g^{2} + g/\tau}\Big),
\end{equation}
which according to Eq. \eqref{1/02-1} for $\hat{M}_{1}(g) = \tau g^{2} + g$  gives $\hat{q}_{1}(\tau; x, g) = \hat{q}_{\rm C}(\tau; x, g)$. That enables us to propose the another kind of the integral decomposition for the generalized Cattaneo equation, that is
\begin{equation}\label{3/02-7}
q_{\hat{\gamma}}(\tau; x, t) = \int_{0}^{\infty} q_{\rm C}(\tau; x, \xi) f_{0}(\xi, t) \D\xi.
\end{equation}
where $f_{0}(\xi, t)$ has been defined in Eq. \eqref{3/03-2}. 

To claim that $f_{0}(\xi, t)$ subordinates $q_{\rm C}(\tau; x, t)$ we should show that the solution of the Cattaneo equation $q_{\rm C}(\tau; x, t)$ is an infinitely divisible PDF. To show that we shall use the connection between CBF and an infinitely divisible PDF presented in Appendix \ref{a2}. The non-negativity as well as the normalizability of $q_{\rm C}(\tau; x, t)$ was extensively studied in Ref. \cite{KGorska20}. Both these properties are ensured by the fact that $[\hat{M}_{1}(z)]^{1/2}$ is CBF which also guarantees that $q_{\rm C}(\tau; x, t)$ is an infinitely divisible.  Thus the solution of the Cattaneo equation $q_{\rm C}(\tau; x, t)$ is the ``safe'' PDF. From the other side the same conclusion can be derived from the central limit theorem if we consider the behaviour of $|x|$ larger than $\mathcal{O}(\sqrt{t})$ \cite{JBKeller04}. The random walk model which corresponds to the Cattaneo and/or generalized Cattaneo equation can be found in, e.g., \cite{KGorska20,MKac74, GHWeiss02, JMasoliver16, JMasoliver17}. 

The solution of the Cattaneo equation $q_{\rm C}(\tau; x, t) = \mathcal{L}^{-1}[\hat{q}_{\rm C}(\tau; x, z); t]$ for general initial conditions is known for a long time, see e.g. \cite[Eq. (102)]{JAStratton41} and \cite[Eq. (7.4.28)]{PMMorse53} which for $q_{\rm C}(\tau; x, 0) = \delta(x)$ and $\dot{q}_{\rm C}(\tau; x, t)|_{t=0} = 0$ reduces to 
\begin{align}\label{3/02-8}
\begin{split}
& q_{\rm C}(\tau; x, t) = \frac{1}{2} \E^{-t/(2\tau)} [\delta(x - at) + \delta(x + at)] \\
& \quad + \Theta\Big(\!t - \frac{|x|}{a}\Big) \frac{\E^{-t/(2\tau)}}{4\tau a} \Big[I_{0}\Big(\frac{1}{2 \tau a} \sqrt{a^{2}t^{2} - x^{2}}\Big) \\
& \quad + \frac{a t}{\sqrt{a^{2} t^{2} - x^{2}}}\, I_{1}\Big(\frac{1}{2 \tau a} \sqrt{a^{2}t^{2} - x^{2}}\Big)\Big],
\end{split}
\end{align}
found in, e.g., \cite{JMasoliver96,HDWeymann67,MAOlivares96}. Notation used in Eq. \eqref{3/02-8} is standard:  the $\delta$-Dirac distribution is denoted as $\delta(\cdot)$,  $\Theta(\cdot)$ means the Heaviside step function and the modified Bessel function of the first kind are marked as $I_{0}(\cdot)$ or $I_{1}(\cdot)$.

Hence, because $q_{\rm C}(\tau; x, t)$ and $f_{0}(\xi,t)$ are ``safe'' PDF's, then Eq. \eqref{3/02-7} can be interpreted as the subordination and from this place we shall use it as a  representation of $q_{\hat{\gamma}}(x, t)$. {We mention that the passage between Eqs. \eqref{2/02-5} and \eqref{3/02-7} exists and relies on the change of functions which contain parameter $\tau$: once it is $f_{\tau}(\xi, t)$ while another time it is $p_{\rm C}(\tau; x, \xi)$. Details of this passage are presented in Appendix \ref{a3a}. } 

\section{Two limits}\label{sec3}

\subsection{The limit $\tau\to 0$:  passage to $N_{\hat{\gamma}}(x, t)$}\label{sec3a}

The passage from $q_{\hat{\gamma}}(\tau; x, t)$ to $N_{\hat{\gamma}}(x, t)$ can be done by demonstrating that for $\tau\to 0$ $q_{\rm C}(\tau; x, t)$ reduces to the normal distribution $N(x, t)$. To show that for $\tau \to 0$ $q_{\rm C}(\tau; x, t)$ tends to $N(x, t)$ let us observe that the term containing $\delta$-Dirac distributions decreases exponentially for small $\tau$ and the Heaviside step function $\Theta$ does not vanish only if $t\in\mathbb{R}_{+}$ and its nonzero value is equal to $1$.  Hence, we conclude that in Eq. \eqref{3/02-8} considered for small $\tau$ only the terms with Bessel functions are essential. Namely, for $\tau \ll 1$ we have
\begin{multline}\label{3/03-1}
q_{\rm C}(\tau; x, t) \propto  \frac{\E^{-t/(2\tau)}}{4\tau a} \Big[I_{0}\big(\ulamek{1}{2\tau a}\sqrt{a^{2} t^{2} - x^{2}}\big) \\ \frac{a t}{\sqrt{a^{2} t^{2} - x^{2}}} I_{1}\big(\ulamek{1}{2\tau a}\sqrt{a^{2} t^{2} - x^{2}}\big)\Big]
\end{multline}
Next, we use the asymptotic formula for $\tau \ll 1$ derived in Appendix \ref{a4}:
\begin{multline}\label{7/02-1}
I_{\nu}\Big(\ulamek{1}{2a \tau}\sqrt{a^{2} t^{2} - x^{2}}\Big) \propto \sqrt{\frac{\tau}{\pi t}}\, \Big(\!1 - \frac{x^{2}}{a^{2}t^{2}}\Big)^{\nu/2} \\
\times \exp\Big(\! - \frac{x^{2}}{4t \tau a^{2}} + \frac{t}{2\tau}\Big), \quad \nu > 0.
\end{multline}
That allows us to reduce Eq. \eqref{3/03-1} to the normal distribution $N(x, t)$. Here we would like to mention that the same result was obtained for $|x| \ll at$ and $a\tau$ small and compared with the mean free path of diffusing objects. In such a case the arguments of modified Bessel functions, as well as these functions themselves, rapidly start to be large. Hence, the first line in Eq. \eqref{3/02-8} containing the $\delta$-Dirac distributions becomes negligible. For the modified Bessel functions once again we use the formula \eqref{7/02-1}. Step by step calculations for $|x| \ll at$ may be found in \cite[Subsec. V.C.]{HDWeymann67}.

Resumming, we can conclude that $\lim_{\tau\to0} q_{\rm C}(\tau; x, t) = \lim_{|x| \ll at} q_{\rm C}(\tau; x, t) = N(x, t)$, Eq. \eqref{3/02-7} reduces to Eq. \eqref{2/02-5} and Eq. \eqref{20/04-1} is reconstructed.

\subsection{The limit $\tau\to \infty$: passage to $W_{\hat{\gamma}}(x, t)$}\label{sec3b}

Let us notice that for large $\tau$ the argument of modified Bessel functions starts to be small such that  in the limit $\tau\to \infty$ the square bracket in Eq. \eqref{3/02-8} is constant. Moreover, it is multiplied by $\exp[-t/(2\tau)]/(4a\tau)$ which vanishes in this limit. The only survivors are  terms with $\delta$-Dirac distributions which we rename as 
\begin{equation}\label{4/03-2}
W(x, t) = \frac{1}{2}[\delta(x - a t) + \delta(x + at)].
\end{equation}
In the above one immediately recognizes non-negative, normalized and infinitely divisible fundamental solution of the wave equation. Thus the integral decomposition \eqref{3/02-7} in the limit of $\tau\to\infty$ reads
\begin{align}\label{4/03-3}
\begin{split}
W_{\hat{\gamma}}(x, t) & = \int_{0}^{\infty} W(x, \xi) f_{0}(\xi, t) \D\xi \\
& = \frac{1}{2 a} [f_{0}(x/a, t) + f_{0}(-x/a, t)].
\end{split}
\end{align}
Because of $W(x, \xi)$ and $f_{0}(\xi, t)$ are "safe" PDFs we conclude that Eq. \eqref{4/03-3} represents the subordination approach.

\section{$f_{0}(\xi, t)$ and $q_{\hat{\gamma}}(x, t)$ -- examples of mutual correspondence}\label{sec4}

\noindent
{\bf (a)} Eq. \eqref{3/02-2} for $\hat{\gamma}(z) = 1$ reads 
\begin{equation*}
f_{0}(\xi, t) = \mathcal{L}^{-1}[\exp(-\xi z); t] = \delta(\xi - t).
\end{equation*}
Substituting it into Eq. \eqref{3/02-7} we obtain  $q_{1}(\tau; x, t) = q_{\rm C}(\tau; x, t)$ supported in the compact domain $\Delta(x,t)$. It is clear that the standard Cattaneo equation is local in time and it does not involve any memory effects. 
\\

\noindent
{\bf (b)} Here we pass to the example which, in contradistinction to {\bf (a)}, involves the memory. As an exactly solvable illustrative model  we take the commonly used power-law memory $\gamma(t) = t^{-\alpha}/\Gamma(1-\alpha)$,   with $\alpha \in (0, 1]$. Its Laplace transform is $\hat{\gamma}(z) = z^{\alpha - 1}$. That gives 
\begin{align}\label{8/02-1}
\begin{split}
f_{0}(\alpha; \xi, t) = \mathcal{L}^{-1}[z^{\alpha - 1} \E^{-\xi z^{\alpha}}; t] = \frac{t}{\alpha \xi}\, g_{\alpha}(\xi, t),
\end{split}
\end{align}
studied e.g., in \cite{KAPenson16, KGorska20}. Two argument one-sided L\'{e}vy stable distribution $g_{\alpha}(\xi, t) = \mathcal{L}^{-1}[\exp(-\xi z^{\alpha}); t]$ can be written with the help of the one argument one-sided L\'{e}vy stable distribution, namely $g_{\alpha}(\xi, t) \equiv \xi^{-1/\alpha} g_{\alpha}(t \xi^{-1/\alpha})$ which expresses the self-similarity property. The explicit form of the inverse Laplace transform in Eq. \eqref{8/02-1}, for $\alpha = l/k$ such that $l < k$ are integers, was found in \cite{KAPenson10, KGorska20a}. Recall that the one-sided L\'{e}vy stable distribution $g_{\alpha}(\sigma)$ is nonzero for $\sigma > 0$ and it vanishes for $\sigma \leq 0$ \cite{KAPenson10}. The nonzero values of $g_{\alpha}(\sigma)$ for $\alpha = l/k$ can be expressed in terms of the Meijer G function $G^{m, n}_{p, q}\big(z\big\vert {(a_{p}) \atop (b_{q})}\big)$ \cite[Eq. (2)]{KAPenson10}. Explicitly this relation reads
\begin{equation}\label{8/03-3}
g_{l/k}(\sigma) = \frac{\sqrt{k l}}{(2\pi)^{(k-l)/2}} \frac{1}{\sigma} G^{k, 0}_{l, k}\Big(\frac{l^{l}}{k^{k} \sigma^{l}}\Big\vert {\Delta(l, 0) \atop \Delta(k, 0)}\Big)
\end{equation}
and appears very handful, both in analytic and numerical calculations, as the Meijer G function is implemented in the computer algebra systems like Mathematica or Maple which makes its use quite friendly. According to common convention the special list of $n$ elements is equal to $\Delta(n, a) = a/n, (a+1)/k, \ldots, (a+n-1)/n$ which is placed as the upper parameters (it is the list $\Delta(l, 0)$) as well as the lower ones (it is the list $\Delta(k, 0)$).

The power-law memory function  $\gamma(t)\propto t^{-\alpha}$, i.e., $\tilde{\gamma}(z)=z^{\alpha-1}$,  allows us to rewrite the generalized Cattaneo equation as the fractional Cattaneo equation with the (Caputo) fractional time derivatives of order $\alpha$, ${_{c}D}_{t}^{\alpha}$. To simplify the notation we denote $q_{z^{\alpha - 1}}(\tau; x, t)$ as $q(\alpha, \tau; x, t)$. Thus the fractional Cattaneo equation takes the form 
\begin{equation}\label{5/03-1}
\tau{_{c}D}_{t}^{2 \alpha} q(\alpha, \tau; x, t) + {_{c} D}_{t}^{\alpha} q(\alpha, \tau; x, t) = B \partial^{2}_{x} q(\alpha, \tau; x, t)
\end{equation}
with $\alpha \in (0, 1]$. For $\alpha = 1$ it looses the fractional character and becomes the standard Cattaneo equation. Observe that $\alpha = 1/2$ separates  Eq. \eqref{5/03-1} into two distinguishable, ``safe'' and ``dangerous'',  cases. For $\alpha\in (0, 1/2]$ the highest order of the Caputo fractional derivative is at most $1$ ( which characterizes the ordinary diffusion equation) and for $\alpha < 1/2$ the order of both derivatives is smaller than $1$. For $\alpha \in (1/2, 1]$ the order of one of the Caputo fractional derivative is larger than $1$ while the order of the second one is smaller than $1$. That leads to different properties, namely the different MSDs and random walks models underlying these equations as it has been discussed in \cite{KGorska20, JMasoliver16}. For $\alpha < 1/2$ and in the limit of small/large $t$ Eq. \eqref{5/03-1} leads to the MSDs charactecteristic for subdiffusion - thus we shall call it the subdiffusion-type equation. Recall that the solution of the standard subdiffusion equation,  e.g. Eq.\eqref{22/03-1} for $\mu<1$, is well known and is presented as the integral decomposition given by Eq. \eqref{8/02-10} \cite{AChechkin21,AStanislavsky19}. In the case of Eq. \eqref{5/03-1} we get  $f_{\tau}(\xi, t)$ instead of $f_{0}(\xi, t)$. Thus, we have Eq. \eqref{2/02-5} from which we can go directly to the integral decomposition \eqref{3/02-7} as may be shown making the calculation presented in Appendix \ref{a3a} for $\hat{\gamma}(z) = z^{\alpha - 1}$.

\begin{figure}[!h]
\includegraphics[scale = 0.36]{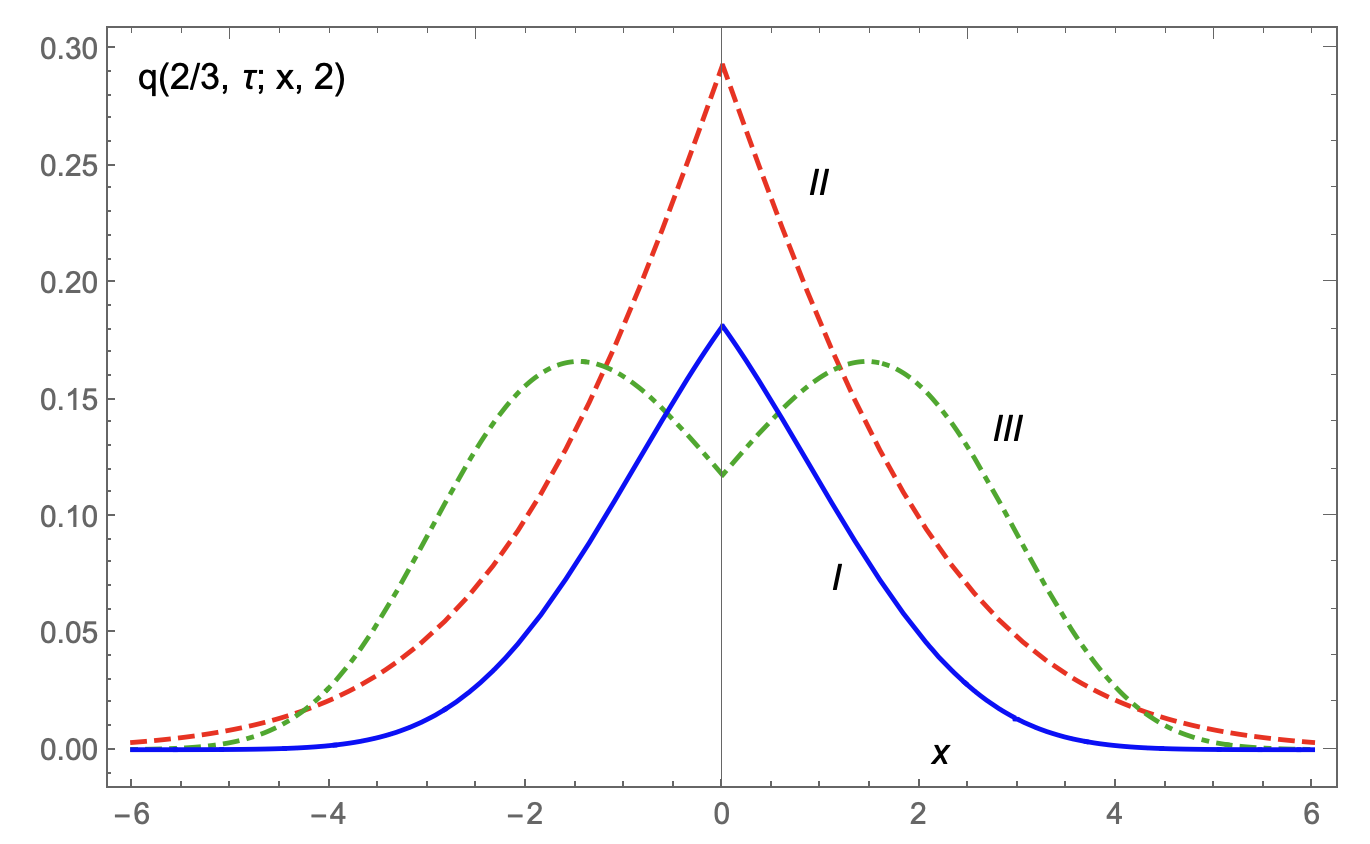}
\caption{\label{fig1} The density $q(2/3, \tau; x, 2)$ for various $\tau$. The red dashed curve (no. II) represents  $\lim_{\tau\to 0} q(2/3, \tau; x, t) = N(2/3; x, 2)$ given by Eq. \eqref{8/04-1} where we set $B = 1$. The green dot-dashed curve (no. III) exhibits $\lim_{\tau\to\infty} q(2/3, \tau; x, 2) = W(2/3; x, 2)$ which is given by Eq. \eqref{12/04-1} with $a = 1$. The blue solid curve (no. I) illustrates the solution of generalized Cattaneo equation \eqref{3/02-7}. The function $f_{0}$ is given by the representation Eq. \eqref{10/03-10}.}
\end{figure}

According to Eq. \eqref{3/02-7} the solution $q(\alpha, \tau; x, t)$ is represented by the integral taken over $\xi\in{\mathbb R}_{+}$, i.e., on the whole positive semi-axis. Restrictions implied by $q_{\rm C}(\tau; x, \xi)$ concern the relation between the laboratory space coordinate $x$ and internal time $\xi$. Such introduced functional dependence between $x$ and $\xi$ is nonzero only on a compact domain but the second term which enters the integrand, namely the subordinator $f_{0}(\xi, t)$, is nonzero for all $\xi, t > 0$. This leads to a new relation between the coordinates $x$ and $t$, by no means sharing the property which $q_{\rm C}(\tau; x, t)$ does obey, i. e., being non-zero only on a compact support.  It means that $q(\alpha, \tau; x, t)$ is nonzero for $x\in\mathbb{R}$. This statement is illustrated in Fig. \ref{fig1} presenting the plot of  $q(\alpha, \tau; x, t)$ as the function of $x$ and $t = 2$. To get the plot shown in Fig. \ref{fig1} we have calculated the integral decomposition given by Eq. \eqref{3/02-7} using the Mathematica 12 software. Performing the calculations we used the function $g_{\alpha}(\sigma)$ expressed by the Meijer G function given by Eq. \eqref{8/03-3} for $\alpha = l/k$, $l, k = 1, 2, \ldots$ which gives 
\begin{equation}\label{10/03-10}
f_{0}(l/k; \xi, t) = \frac{k^{3/2} l^{-1/2}}{(2\pi)^{(k-l)/2}} \frac{1}{\xi} G^{k, 0}_{l, k}\Big(\frac{l^{l} \xi^{k}}{k^{k} t^{l}}\Big\vert {\Delta(l, 0) \atop \Delta(k, 0)}\Big).
\end{equation}

In Fig. \ref{fig1} we present also the special cases of $q(\alpha, \tau; x, t)$ obtained by taking its limits of $\tau\to 0$ and $\tau\to\infty$. For $\tau\ll 0$ we can use Eq. \eqref{8/02-10} in which $N(x, \xi) = (2 \sqrt{B})^{-1} f_{0}(1/2; |x|/\sqrt{B}, \xi)$. Then, we employ \cite[Eqs. (29)-(32)]{KAPenson16}. That allows one to find
\begin{equation}\label{8/04-1}
N(\alpha; x, t) = (2 \sqrt{B})^{-1} f_{0}(\alpha/2; |x|/\sqrt{B}, t),
\end{equation}
equal to \cite[Eq. (16)]{EBarkai01}. Eq. \eqref{8/04-1} can be also obtained by using the technique presented in  \cite{KGorska12, EBarkai01}. For $\tau\gg 1$, with $a=(B/{\tau})^{1/2}$, \ Eq. \eqref{4/03-3} reads
\begin{equation}\label{12/04-1}
W(\alpha; x, t) = (2a)^{-1} f_{0}(\alpha; |x|/a, t).
\end{equation}
For $a=1$ it is equivalent, e.g., to \cite[Eq. (9)]{YuLuchko13} or \cite[Eq. (33)]{YuLuchko19}, which employ the Mainardi function $M_{\alpha}(u) = u^{-1-1/\alpha} g_{\alpha}(u^{-1/\alpha})/\alpha$. Eqs. \eqref{8/04-1} and \eqref{12/04-1}, although at the first glance look quite similar, in fact  differ significantly. The difference, clearly seen in Fig. \ref{fig1}, has its origin in the value of the first parameter which is either $\alpha/2$ in Eq. \eqref{8/04-1} or $\alpha$ in Eq. \eqref{12/04-1}. For the solutions to the (standard) anomalous diffusion equation given by Eq. \eqref{8/04-1} we always observe one maximum localized at $x = 0$ while for the diffusion-wave equation we get either one or two maxima. The number of maxima in question depends on the value which the parameter $\alpha$ takes on: for $\alpha\in(0, 1/2]$ we deal with one maximum localized at $x = 0$ while for $\alpha\in(1/2, 1]$ we have two maxima localized at $x = \pm x_{m}$ (the curve no. III in Fig \ref{fig1}; $x_{m} = 1.463$). Here we remind that such behaviour was noticed and discussed e.g. in \cite{YFujita90, YuLuchko13} in the context of propagation velocity characterizing diffusion phenomena.\\

\begin{figure}[!h]
\includegraphics[scale = 0.36]{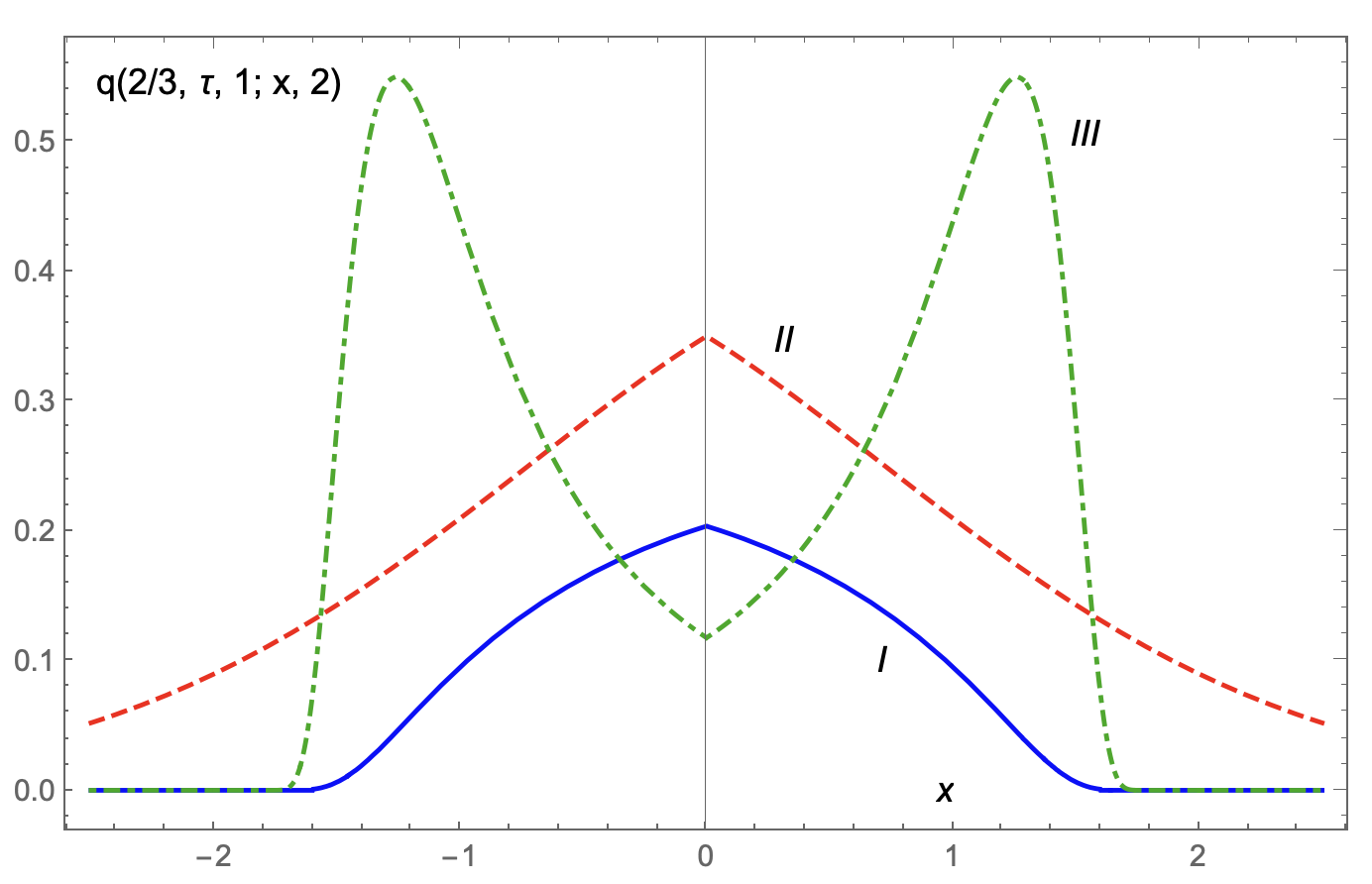}
\caption{\label{fig2} The figure represents the PDF $q(2/3, \tau, 1; x, 2)$ (the blue solid curve, no. I) and its limits with respect to $\tau$. For small $\tau$ it goes to $N(2/3, 1; x, 2)$ (the red dashed curve, no. II) and for large $\tau$ it tends to $W(2/3, 1; x, 2)$ (the green dot-dashed curve, no. III). The function $f_{0}$ is calculated using Eq. \eqref{5/03-2} in which the one-sided L\'{e}vy stable distribution $g_{l/k}(\sigma)$, $0 < l < k$, is given by Eq. \eqref{8/03-3}.}
\end{figure}

\noindent
{\bf (c)} The solution $q_{1}(\tau; x, t)$ of the example {\bf (a)} is defined on the compact support. This situation becomes opposite, as seen from {\bf (b)}, when the memory effects, in fact meaning the time non-locality, come to play.  To investigate this change we will consider the example {\bf (b)} perturbed by the non-negative constant parameter $\epsilon$, i.e., a linear combination of memories which define examples {\bf(a)} and {\bf (b)}. Such new memory has the form $\hat{\gamma}(z) = z^{\alpha-1} + \epsilon$ where $\alpha\in(0, 1]$ and $\epsilon \geq 0$. That leads to
\begin{align}\label{5/03-2}
f_{0}(\alpha, \epsilon; &\, \xi, t) = \mathcal{L}^{-1}[(z^{\alpha-1} + \epsilon)\E^{-\xi z^{\alpha}} \E^{-\xi\epsilon z}; t] \nonumber\\
& = \Theta(t-\epsilon\xi) \left(\frac{t}{\alpha\xi} - \epsilon\frac{1-\alpha}{\alpha}\right) g_{\alpha}(\xi, t-\epsilon \xi), 
\end{align}
where the inverse Laplace transforms have been calculated applying twice the Laplace convolutions once for $z^{\alpha - 1} \exp(-\xi z^{\alpha})$ and another time for $\exp(-\xi z^{\alpha})$. The occurrence of $\Theta(t-\epsilon\xi)$ guarantees that the second argument of $g_{\alpha}(\xi, t-\epsilon\xi)$ is non-negative. Consequently, it changes the infinite range integral given by Eq. \eqref{3/02-7} into the integral over the finite range. Thus, we get
\begin{multline}\label{8/03-5}
q_{z^{\alpha-1} + \epsilon}(\tau; x, t) \equiv q(\alpha, \tau, \epsilon; x, t) = \int_{0}^{t/\epsilon}\!\! q_{\rm C}(\tau; x, \xi)\\ \times\left(\frac{t}{\alpha\epsilon\xi} - \frac{1-\alpha}{\alpha}\right) g_{\alpha}(\xi, t-\epsilon \xi)\!\D\xi.
\end{multline}
Because this integral is defined in the finite region and $q_{\rm C}(\tau; x, \xi)$ contains the $\delta$-Dirac distributions as well as the Heaviside step function we get the relation between $x$ and $t$ in which $|x| \leq a t$ for $\epsilon = 1$. Hence, we can suspect that $q(\alpha, \tau, \epsilon; x, t)$ for $\epsilon > 0$ is nonzero only in the finite subset of $x$. The size of this subset enlarges as the evolution goes on and its dimension depends on the time $t$ passed, as well as on the constants $a$ and $\epsilon$. The above relation between $x$ and $t$ disappears for $\epsilon\to 0$ and  $q(\alpha, \tau, 0; x, t)$  becomes nonzero for all $x\in\mathbb{R}$, i.e., we arrive at the example {\bf (b)}. Analytic calculations of Eq. \eqref{8/03-5} are difficult to be done so we present only numerical results obtained using Mathematica 12. Results are plotted on Fig. \ref{fig2} obtained for rational $\alpha = l/k$ and Eqs. \eqref{3/02-8} and \eqref{8/03-3} employed. 

In Fig. \ref{fig2}, besdes of the example of $q(\alpha, \tau, \epsilon; x, t)$ for $\epsilon > 0$ (the blue curve,  no. I), there are also presented its two special cases in which we consider the limits of $q(\alpha, \tau, \epsilon; x, t)$ for $\tau\to 0$ and $\tau\to\infty$. In the first case the PDF $q(\alpha, 0, \epsilon; x, t)$ is equal to Eq. \eqref{8/03-5} where instead of $q_{\rm C}(\tau; x, \xi)$ sits $N(x, \xi)$. This limit is denoted as $N(\alpha, \epsilon; x, t)$ and it is presented in Fig. \ref{fig2} by the red curve no. II. In the next limit we have $q(\alpha, \infty, \epsilon; x, t)$ denoted as $W(\alpha, \epsilon; x, t)$. It is given though Eq. \eqref{4/03-3} in which $f_{0}$ is equal to Eq. \eqref{5/03-2} . The PDF $W(\alpha, \epsilon; x, t)$ is illustrated by the green curve with no. III.

\section{Discussion}\label{sec5}

The main results of the paper may be resumed  two-fold. Firstly, we shed new light on  the subordination approach and present its modification naturally appearing in the description of anomalous diffusion. Secondly, we provide an explicit example of the memory functions ${\hat{\gamma}}(s)$ which if put into generalized, time smeared, Cattaneo equation lead to solutions sharing the basic property obeyed by solutions of the standard Cattaneo equation, namely it do vanish outside the compact region $\Delta(x,t)$, i.e. $q_{\hat{\gamma}}(\tau; x, t)=0$ if $x,t\notin\Delta(x,t)$.

The integral decompositions of solutions to the diffusion-like equations are rooted in procedures which allow to transform solutions found in the Fourier/Laplace domain to the physical $(x,t)$ space. If merged with principles of the subordination approach the integral decompositions serve as a tool which makes it possible to prove the probabilistic character of such obtained solutions. However, as has been warned in recent papers, this claim has to be treated carefully. The reason is that integral decompositions constructed according to the standard procedures are not always sufficient to confirm the probabilistic interpretation of $q_{\hat{\gamma}}(\tau; x, t)$ in a complete and correct way \cite{AChechkin21,IMSokolov05,TSandev18}. The commonly proposed integral decompositions are assumed to reflect the fact that the Brownian motion, described by the normal distribution $N(x, \xi)$  parametrized in physical space $x$ and operational time $\xi$, is subordinated by a random process which links the physical and operational times through the relation given by an infinite divisible distribution $f_{\tau}(\xi, t)$. Despite the fact that the normal distribution $N(x, \xi)$ satisfies all conditions needed to be a suitable PDF there exist examples for which the second component of such a way obtained decomposition, namely the function $f_{\tau}(\xi, t)$, is not the PDF \cite{TSandev18}. In our paper we propose another type of integral decomposition within which, instead of the normal distribution $N(x, t)$, there appears  the fundamental solution of the standard Cattaneo equation $q_{\rm C}(\tau; x, t)$. The latter satisfies all requirements to be a suitable PDF  and using it allows us to keep the subordinator $f_{0}(\xi, t)$ which usually accompanies $N(x, t)$. Correctness and validity of such introduced modification of the subordination approach is confirmed by showing that in the limits of small and large $\tau$ the PDF $q_{\hat{\gamma}}(\tau; x, t)$ reconstructs solutions $N_{\hat{\gamma}}(x, t)$ and $W_{\hat{\gamma}}(x, t)$ of the time smeared diffusion and wave equations.

The second important result flows out from the analysis of three examples presented in the paper and illustrating the role played by memory functions. Our examples allow us to suspect which shape the memory functions should have in order to judge if the solution of generalized Cattaneo equation $p_{\hat{\gamma}}(\tau; x, t)$ has non-zero values only in the compact region and vanishes outside of it. Thus, we propose the answer on the open problem about vanishing of the solution of the anomalous diffusion and/or diffusion-wave equations. As the first example we consider the standard Catteneo equation  in which the second and first time derivatives are strictly localized. Solution to this equation $q_{1}(\tau; x, t) \equiv q_{\rm C}(\tau; x, t)$  is nonzero for any $x\in\Delta(x,t)$. The situation changes radically if we introduce the memory functions $\eta(t)$ and $\gamma(t)$. For the predominantly used model of memory functions, i.e. $\eta \propto t^{1-2\alpha}$ and $\gamma(t) \propto t^{-\alpha}$, $\alpha\in(0, 1]$, the PDF $q_{z^{\alpha - 1}}(\tau; x, t) = q(\alpha, \tau; x, t)$ is always nonzero for $x\in\mathbb{R}$. Analogical situation takes place for both limit cases, $\tau\to 0$ and $\tau\to\infty$  - the solutions $N(\alpha; x, t)$ as well as $W(\alpha; x, t)$ are nonzero for all $x\in\mathbb{R}$. Different behaviour we face if consider the second example perturbed by a strictly localized term - this we model by introducing memories $\eta(t) = \epsilon\delta(t) + t^{1-2\alpha}/\Gamma(2 - 2\alpha)$ and $\gamma(t) = \epsilon \delta(t) + t^{-\alpha}/\Gamma(1-\alpha)$, $\epsilon \geq 0$, anytime with $\alpha\in(0, 1]$. Obviously, putting $\epsilon = 0$ we come back to the second example and to the generalized Cattaneo equation which solution is nonzero for $x\in\mathbb{R}$. For $\epsilon > 0$ the solution behaves  differently: $q_{z^{\alpha-1} + \epsilon}(\tau;x,t) \equiv q(\alpha, \tau, \epsilon; x, t)$ does not vanish only for $x \in \Delta(x,t/\epsilon)$. Such compact region appears in the limit of $\tau\to\infty$ for which $q(\alpha, \infty, \epsilon; x, t) = W(\alpha, \epsilon; x, t)$ but does not happen for $\tau\to 0$ in which $q(\alpha, 0, \epsilon; x, t)$ equals $N(\alpha, \epsilon; x, t)$. 

Sec. \ref{sec3a} extends results of the seminal paper \cite{HDWeymann67}. It shows that for $|x| \ll at$ it is not possible to differentiate between the plots of solutions to standard anomalous diffusion \cite{RMetzler19,YuLuchko19} and generalized Cattaneo equations \cite{KGorska20,EAwad20,JMasoliver17,JMasoliver16,ACompte97}. Having this at hand one cannot show which of patterns fits the experimental data better. Thus, if we are limited to just mentioned range of $x$, we are unable to decide which random walk model, either continuous time random walk (CTRW) or, alternatively, continuous time persistent random walk (CTPRW) is more appropriate to describe phenomena under consideration. These models differ by the probability densities. In CTRW the random walker arrives to the localization $(x, t)$ moving along one channel whereas in CTPRW it may arrive to $(x, t)$ through two channels. This dichotomy results in the differences of waiting time and jump PDFs. In CTRW the waiting time PDF in the Laplace domain reads $\hat{\psi}(s) = [1 + \tau s^{2} \hat{\gamma}^{2}(s) + s \hat{\gamma}(s)]$ whereas in CTPRW we have $\hat{\psi}_{\pm} = [1 + 2\tau s \hat{\gamma}(s)]^{-1}$ \cite{KGorska20}. The CTRW and CTPRW models differ also by the jump PDFs, in CTRW it depends only on the position $x$ whereas for CTPRW it depends both on the position $x$ and time $t$.

\section{Conclusions}\label{sec6} 

In our opinion observations made in Sec. \ref{sec4} and Disscusion justify to formulate the conjecture: 
\begin{quote}
{\em This is the appearance of the $\delta$-Dirac distribution in memory function $\eta(t)$ (so involving the constituent leading to  strictly localized second time derivative) which ensures that the generalized Cattaneo equation can be interpreted in terms of the wave equation. In such a case  its solutions share properties characteristic for the wave-like propagation and are nonzero only in a compact region defined by motion of the wave fronts.} 
\end{quote}
The above expresses the claim that the second time derivative present in the evolution equation prevents any particle to spread out beyond the compact region similarly  like it takes place for the Cattaneo equation \cite{SGoldstein51}. We are going to verify this conjecture by checking for which forms of memory function (if any besides of one discussed in Sec. \ref{sec3}A) the wave behavior of $p_{\hat{\gamma}}(\tau; x, t)$ may be observed. We expect that further analysis of wave phenomena seen in solutions to the Cattaneo equation, like the Doppler studied recently in Ref. \cite{YuPovstenko21}, will appear helpful to understand better wave properties of solutions to the generalized Cattaneo equation.

Figs. \ref{fig1} and \ref{fig2} illustrate the cusp-containing and bimodal behavior (occurring in dependence to the value of $\tau$) of solutions relevant for examples {\bf (b)} and {\bf (c)} of Sec. \ref{sec4}. Such patterns have been noted in semi-relativistic time evolution equations investigated in classical and quantum mechanics, to mention the Salpeter equation \cite{CBeck03,KKowalski11,KAPenson18,MVChubynsky14,RJain17}. In studies of anomalous diffusion cusp-containing and bimodal behavior of PDF-interpreted solutions have appeared useful and quite popular in considerations  dealing with the so-called Brownian but non-Gaussian models \cite{SHapca09,BWang09a,BWang12,AVChechkin17,WWang20,WWang20a,RMetzler20}. Our approach suggests possible modifications of results presented in \cite{MVChubynsky14, AVChechkin17, VSposini18, EBPostnikov20} devoted to the concept of diffusing diffusivity model being recently a hot topic in anomalous diffusion research stimulated by tracking of single particle diffusion. In diffusing diffusivity models the diffusion coefficient $B$ of the tracer particle is supposed to evolve in time and its time behaviour is (usually) assumed to take the shape which the coordinate of a Brownian particle in a gravitational field has. In such a case the distribution function $P(x, t)$ of a system of traced particles is given by 
\begin{equation}\label{1/07-1}
P(x, t) = \int_{0}^{\infty} N(x, t| B) p_{B}(B) \D B,
\end{equation}
where $N(x, t| B) \equiv N(x, t)$ is used to denote explicitly the dependence on  $B$. The PDF $p_{B}(B)$ is usually assumed to be $\exp(-B/\langle B \rangle)/\langle B \rangle$ with $\langle B \rangle$ being the effective diffusivity. Hence, $P(x, t)$ is given by the Laplace distribution, see \cite{SHapca09,BWang09a,AVChechkin17}. Currently investigated modifications of Eq. \eqref{1/07-1}, discussed in \cite{AVChechkin17,VSposini18,AGCherstvy21}, rely on changing $p_{B}(B)$ but keeping $N(x, t| B)$. Based on results presented in our paper we would like to propose another modification of Eq. \eqref{1/07-1}, namely to substitute $p_{\hat{\gamma}}(\tau; x, t)$ instead of $N(x, t| B)$ but still preserve the exponential form of $p_{B}(B)$. This will lead to the cusp-containing and bimodal behaviour which appropriately chosen memory function may cause that diffusion will be restricted to the compact region.

\section*{Acknowledgments}

K.G. expresses her gratitude to the anonymous referees whose critical remarks and a numerous  suggestions help her very much to improve the paper such physically like mathematically. Especially, I appreciate it very much the first referee for suggesting me to get interested in the diffusive diffusivity concept. 

The author thanks Prof. Eli Barkai for asking her the question concerning the region for which the solution of the generalized Cattaneo equation is defined. I believe that this paper gives answer his question. I am also very grateful to Prof. Andrzej Horzela for long discussions on the subject.

The research was supported by the NCN (National Research Center, Poland) Research Grant OPUS-12 No. UMO-2016/23/B/ST3/01714 as well as by the NCN (National Research Center, Poland) and the NAWA (Polish National Agency For Academic Exchange, Poland) Research Grant Preludium Bis 2 No. UMO-2020/39/O/ST2/01563. 

\appendix
\section{The completely monotonic and completely Bernstein functions}\label{a1}
\noindent
The function $\widehat{h}(s)$, $s\in\mathbb{R}_{+}$, is a completely Bernstein function (CBF) if: 
\begin{itemize}
\item[1.] $\widehat{h}(s)$ and its derivative $[\widehat{h}(s)]'$ are non-negative functions and, moreover, all derivatives of $[\widehat{h}(s)]'$ alternate \cite{RLSchilling10}, namely
\begin{equation*}
(-1)^{n-1} \widehat{h}^{(n)}(s) \geq 0, \quad n = 1, 2, \ldots;
\end{equation*}
\item[2.] $\widehat{h}(s)$ and $s/\widehat{h}(s)$ have the representation given by the Stieltjes transform \cite{RLSchilling10}.
\end{itemize}
Alternative criterion says that $\widehat{h}(s)$ is CBF if $s/\widehat{h}(s)$ is CBF \cite{RLSchilling10}. \\

\noindent
{The completely monotonic function (CMF)} $\widehat{H}(s)$ are non-negative function of a non-negative argument whose all derivatives exist and alternate, i.e.
\begin{equation*}
(-1)^{n} \widehat{H}^{(n)}(s) \geq 0, \quad n=0, 1, \ldots.
\end{equation*}

{The Bernstein theorem} uniquely and mutually connects CMF with the non-negative function defined on $\mathbb{R}_{+}$ by the Laplace transform:
\begin{equation*}
\widehat{H}(s) = \int_{0}^{\infty} \E^{-s t} H(t) \D t,
\end{equation*}
where $\widehat{H}(s)$ is CMF \cite{DVWidder46, HPollard44, ANKochubei11}.

\section{The relation between the infinitely divisible distribution and the Bernstein-class functions: BF as well as CBF}\label{a2}

The relation between the CBF and infinitely divisible function is expressed by \cite[Lemma 9.2]{RLSchilling10}. It say that the measure $g$ on $[0, \infty)$ is infinitely divisible iff $\mathcal{L}[g; \lambda] = \exp[-f(\lambda)]$ where $f$ is CBF.

\section{The Efross theorem}\label{a3}

We quote here the  Efross theorem \cite{AEfross35,LWlodarski52,UGraf04,KGorska12a,AApelblat21} which is the generalization of the convolution theorem for the Laplace transform. According to it if $G(z)$ and $g(z)$ are analytic functions, ${\cal L}[h_{1}(x, \xi); z] = \hat{h}_{1}(x, z)$ and
\begin{equation*}
{\cal L}[h_{2}(\xi, t); z] = \int_{0}^{\infty} h_{2}(\xi, t) \E^{-z t} \D t = \widehat{G}(z) \E^{-\xi \hat{k}(z)},
\end{equation*}
then
\begin{equation*}
\widehat{G}(z)\hat{h}_{1}(x,\hat{k}(z)) = \int_{0}^{\infty}\!\Big[\!\int_{0}^{\infty}\!\! h_{1}(x, \xi) h_{2}(\xi, t) \D\xi\Big] \E^{- z t} \D t.
\end{equation*}
Thus, we can conclude that
\begin{equation}\label{30/03-4}
{\cal L}^{-1}[\widehat{G}(z)\hat{h}_{1}(x,\hat{k}(z)); t] = \int_{0}^{\infty}\!\! h_{1}(x, \xi) h_{2}(\xi, t) \D\xi.
\end{equation}

We illustrate this theorem by deriving Eq. \eqref{3/02-7}. As $\hat{h}_{1}(x,z)$ we take $F(x,z)$ from Eq. \eqref{23/02-5}. That gives $h_{1}(x, \xi) = {\cal L}^{-1}[F(x,z); \xi]$. This inverse Laplace transform can be calculated by using the formulas placed in \cite[Appendix A]{JMasoliver19} and/or \cite[Appendix B]{YuPovstenko21}. Namely,
\begin{multline}\label{30/03-3}
{\cal L}^{-1}[(z^{2} - \lambda^{2})^{-1/2} \exp[- \ulamek{|x|}{a} (z^{2} - \lambda^{2})^{1/2}]; \xi\} \\
= \Theta(\xi - \ulamek{|x|}{a}) I_{0}[\ulamek{1}{2\tau}(\xi^{2} - \ulamek{x^{2}}{a^{2}})^{1/2}]
\end{multline}
and
\begin{multline}\label{30/03-5}
{\cal L}^{-1}[z (z^{2} - \lambda^{2})^{-1/2} \exp[- \ulamek{|x|}{a} (z^{2} - \lambda^{2})^{1/2}]; \xi\} \\
= \delta(\xi - \ulamek{|x|}{a}) + \Theta(\xi - \ulamek{|x|}{a}) \frac{\lambda\xi}{(\xi^{2} - \ulamek{x^{2}}{a^{2}})^{1/2}} \\ \times I_{1}[\ulamek{1}{2\tau}(\xi^{2} - \ulamek{x^{2}}{a^{2}})^{1/2}],
\end{multline}
where $\lambda = (2\tau)^{-1}$. Due to the property of $\delta$-Dirac distribution we have that $\delta(\xi - |x|/a) = a \delta(a \xi - |x|)$ gives two peaks at $x -a\xi$ and $x + a\xi$. Thus, it can be expressed as $a[\delta(x - a \xi) + \delta(x + a \xi)]$ and 
\begin{equation}\label{30/03-6}
h_{1}(x, \xi) = \E^{\xi/(2\tau)} q_{\rm C}(x, \xi).
\end{equation}
According to Subsec. \ref{ssec2B} we have that $\widehat{G}(z) = \hat{\gamma}(z)$ and $\hat{k}(z) = g(z)$ given by Eq. \eqref{23/02-4}. That leads to  
\begin{align}\label{30/03-7}
\begin{split}
h_{2}(\xi, t) & = {\cal L}^{-1}[\hat{\gamma}(z) \E^{-\xi z \hat{\gamma}(z) - \xi/(2\tau)}; t] \\
& = \E^{-\xi/(2\tau)} f_{0}(\xi, t),
\end{split}
\end{align}
where we used the property of inverse Laplace transform saying that 
\begin{equation}\label{20/04-3}
{\cal L}^{-1}[\hat{H}(z + \lambda); t] = \E^{-t\lambda} {\cal L}^{-1}[\hat{H}(z); t].
\end{equation}

Inserting Eqs. \eqref{30/03-6} and \eqref{30/03-7} into Eq. \eqref{30/03-4} we derive $q_{\hat{\gamma}}(\tau; x, t)$.

\section{The passage between Eqs. \eqref{2/02-5} and \eqref{3/02-7}}\label{a3a}

Eq. \eqref{2/02-5} in which we insert Eq. \eqref{3/02-1} where $\hat{M}_{\hat{\gamma}}(z)$ leads to Eq. \eqref{23/02-2}. If we separate $z \hat{\gamma}(z)$ from $\hat{M}_{\hat{\gamma}}(z)$ then we can express $f_{\tau}(\xi, t)$ as
\begin{equation*}
f_{\tau}(\xi, t) = \mathcal{L}^{-1}\{\hat{\gamma}(z) [1 + \tau z \hat{\gamma}(z)] \E^{-\xi z \hat{\gamma}(z)[1 + \tau z \hat{\gamma}(z)]}; t\}.
\end{equation*}
Using the Efross theorem in which $\hat{k}(z) = z \hat{\gamma}(z)$, $\hat{G}(z) = \hat{\gamma}(z)$, and $\hat{h}_{1}(\xi, y) = (1+ \tau y)\exp[-\xi y(1 + \tau y)]$, we represent $f_{\tau}(\xi, t) $ in the form
\begin{multline}\label{23/06-2}
f_{\tau}(\xi, t) = \int_{0}^{\infty}\!\! \mathcal{L}^{-1}[(1+ \tau y)\E^{-\xi y(1 + \tau y)}; u] \D u \\ \times \mathcal{L}^{-1}[\hat{\gamma}(z) \E^{-u z \hat{\gamma}(z)}; t].
\end{multline}
Then, we rewrite Eq. \eqref{2/02-5} as
\begin{align}\label{23/06-3}
q_{\hat{\gamma}}(&\tau; x, t)  = \int_{0}^{\infty}\!\! \mathcal{L}^{-1}[(1 + \tau y) \nonumber \\ & \times\int_{0}^{\infty}\!\! N(x, \xi) \E^{-\xi y (1 + \tau y)} \D\xi; u]\, f_{0}(u, t) \D u,
\end{align}
where the integral over $\xi$ goes inside the first inverse Laplace transform in Eq. \eqref{23/06-2}. Due to \cite[Eq. (2.3.16.2)]{APPrudnikov-v1} we get $\int_{0}^{\infty}\!\! \E^{-a/\xi - b \xi} \xi^{-1/2} \D\xi = \sqrt{\pi/b} \E^{-2\sqrt{a b}}$. This formula allows one to rewrite Eq. \eqref{23/06-3} as
\begin{equation}\label{24/06-1}
q_{\hat{\gamma}}(\tau; x, t) = \int_{0}^{\infty}\!\! \mathcal{L}^{-1}\{F[x, y + 1/(2\tau)]; u\} f_{0}(u, t) \D u,
\end{equation} 
where $F(x, g)$ is given by Eq. \eqref{23/02-5}. Thereafter, we can repeat calculation in Eq. \eqref{3/02-5} in which instead of $z \hat{\gamma}(z)$ we have $y$ and we use property Eq. \eqref{20/04-3}. At the consequence of it we obtain the integral decomposition given by Eq. \eqref{3/02-7}.

\section{Derivation of Eq. \eqref{7/02-1}}\label{a4}

We start with the series definition of modified Bessel function of the first kind. That allows us to write
\begin{multline}\label{31/03-2}
I_{\nu}(\ulamek{1}{2a\tau}\sqrt{a^{2}t^{2} - x^{2}}) \\ = \sum_{n\geq 0} \frac{[t/(4\tau)]^{2n + \nu}}{n! \Gamma(\nu + n + 1)} \Big(1 - \frac{x^{2}}{a^{2} t^{2}}\Big)^{n + \frac{\nu}{2}}.
\end{multline}
Applying now the binomial sum to $\{1 - [x/(a t)]^{2}\}^{n}$ we can express the modified Bessel function in Eq. \eqref{31/03-2} as
\begin{equation}\label{31/03-3}
\Big(1 - \frac{x^{2}}{a^{2} t^{2}}\Big)^{\nu/2} \sum_{n\geq 0} \sum_{r=0}^{n} \frac{[t/(4\tau)]^{2n + \nu}}{\Gamma(\nu + n + 1)} \frac{(-\ulamek{x^{2}}{a^{2} t^{2}})^{n-r}}{r! (n - r)!}. 
\end{equation}
Then, we observe that the double sum $\sum_{n\geq 0} \sum_{r=0}^{n}$ can be rewritten in the form  $\sum_{r \geq 0} \sum_{n=r}^{\infty}$. Setting $n - r = j$ we have 
\begin{multline}\label{31/03-4}
\Big(1 - \frac{x^{2}}{a^{2} t^{2}}\Big)^{\!\nu/2} \sum_{j\geq 0} \frac{1}{j!} \Big(\!-\frac{x^{2}}{4 \tau a^{2} t}\Big)^{\!j} \\ \times 
\sum_{r\geq 0} \frac{[t/(4\tau)]^{2r + \nu + j}}{\Gamma(\nu + j + r + 1) r!} 
\end{multline}
The series definition of the modified Bessel function $I_{\nu}(\cdot)$ allow us to write
\begin{multline*}
I_{\nu}(\ulamek{1}{2a\tau}\sqrt{a^{2}t^{2} - x^{2}}) \\= \Big(1 - \frac{x^{2}}{a^{2} t^{2}}\Big)^{\!\nu/2} \sum_{j\geq 0} \frac{1}{j!} \Big(\!-\frac{x^{2}}{4 \tau a^{2} t}\Big)^{\!j} I_{\nu + j}\Big(\frac{t}{2\tau}\Big),
\end{multline*}
which agree with \cite[Eq. (5.8.3.1)]{APPrudnikov-v2}. Using the asymptotic form of modified Bessel function for large $t/(2\tau)$, which boils down to $I_{\nu}[t/(2\tau)] = [\tau/(\pi t)]^{1/2} \exp[t/(2\tau)]$, we obtain Eq. \eqref{7/02-1}.

\end{document}